\documentclass{rsproca_new}

\usepackage{graphicx}
\usepackage{hyperref}
\usepackage{color}

\renewcommand{\dblfloatsep}{100}

\newcommand{\indhel}{s}
\newcommand{\indlmf}{\ell mc}

\newcommand{\an}[1]{a_{#1}}
\newcommand{\ank}[2]{a_{#1,#2}}

\newcommand{\As}{{}_{\indhel}\bar{A}_{\indlmf}}
\newcommand{\A}{{}_{\indhel}A_{\indlmf}}
\newcommand{\Ask}[1]{\bar{A}_{#1}}
\newcommand{\E}{{}_{\indhel}E_{\indlmf}}
\newcommand{\Ecz}{{}_{\indhel}E_{\ell m0}}

\newcommand{\q}{{}_{\indhel}q_{\ell m}}

\newcommand{\g}{g}

\jname{rspa}
\Journal{Proc R Soc A\ }

\begin{document}

\title{High-order asymptotics for the Spin-Weighted Spheroidal Equation at large real frequency}

\author{Marc Casals $^{1,2}$, Adrian C.~Ottewill$^{2}$ and Niels Warburton$^{2}$}
\address{
$^{1}$Centro Brasileiro de Pesquisas F\'isicas (CBPF), Rio de Janeiro, CEP 22290-180, Brazil.\\
$^{2}$School of Mathematics and Statistics, University College Dublin, Belfield, Dublin 4, Ireland.
}

\subject{mathematical physics}

\keywords{spin-weighted spheroidal harmonics, angular eigenvalues, black holes}

\corres{Niels Warburton\\
\email{niels.warburton@ucd.ie}}

\begin{abstract}
The spin-weighted spheroidal eigenvalues and eigen-functions arise in the separation by variables of spin-field perturbations of Kerr black holes.
We derive a large, real-frequency asymptotic expansion of the spin-weighted spheroidal eigenvalues and eigenfunctions to high order.
This expansion corrects and extends existing results in the literature and 
we validate it via a high-precision numerical calculation.
\end{abstract}

\maketitle

\section{Introduction} \label{sec:Introduction}

Teukolsky derived a single ``master" equation for spin-field perturbations of rotating 
(Kerr) black  holes~\cite{teukolsky1972rotating,Teukolsky:1973ha}.
This $(3+1)$-dimensional master equation separates by variables, with the polar-angular factor 
in the solution
being
the so-called spin-weighted spheroidal eigenfunction.
The corresponding
eigenvalue also appears in the equation satisfied by the radial factor of the
Teukolsky master solution.
Thus, both the eigenvalues and the eigenfunctions are  important for studying perturbations of astrophysical black holes.

Neither the spin-weighted spheroidal eigenfunctions nor their eigenvalues are known in closed form but they can be calculated using numerical and
analytical techniques (see~\cite{Berti:2005gp,berti2006erratum} for a review).
As for the analytical techniques,
for example, expansions have been   obtained for small frequency~\cite{press1973perturbations,Starobinskil:1974nkd, fackerell1977spin,seidel1989comment}
and 
asymptotic analyses have been carried  out for
large, purely-imaginary frequency~\cite{breuer1975gravitational,berti2004highly,Berti:2005gp,berti2006erratum,hod2013black}.

In this paper we are instead interested in the 
asymptotics for large, {\it real} frequency.
These asymptotics are interesting for various reasons, such as for synchrotron radiation~\cite{chrzanowski1974geodesic,breuer1975gravitational}, for the study of  divergences 
in either the quantum or classical field theories (e.g.,~\cite{CDOWb,PhysRevD.51.4337}
for WKB in the case of  spherically-symmetric space-times and~\cite{Ottewill:2000qh,Casals:Ottewill:2005} for expressions for expectation values
 involving (spin-weighted) spheroidal harmonics in Kerr) and gravitational waves from rapidly rotating black holes \cite{Yang:2013uba,Gralla:2015rpa}. Analytic approximations are also extremely valuable as checks on numerical calculation schemes.
The large, real frequency behaviour of the {\it scalar} spheroidal eigenfunctions and eigenvalues was studied in~\cite{Erdelyi:1953,Flammer,meixner2013mathieusche}.
The first large, real frequency study in the non-zero spin case was carried out in~\cite{breuer1975gravitational}.
However, this work contained an error which was corrected by Breuer, Ryan and Waller~\cite{breuer1977some} (BRW).
BRW provided an asymptotic expansion for the eigenvalue up to six leading orders, which
depended crucially on a parameter $\q$
that was left undetermined for the case of non-zero spin.
Furthermore, their analysis for non-zero spin had an error in the asymptotic behaviour
 of the eigenfunctions which was later corrected in~\cite{Casals:2004zq}.
This correction further allowed~\cite{Casals:2004zq} to  analytically obtain the parameter $\q$ 
as well as the correct first term in a large real-frequency series expansion for the eigenfunctions.

As it turns out, however, the last three orders in the asymptotic expansion for the eigenvalue provided in BRW formally in terms of $\q$ were also incorrect.
In this paper, we correct these 3rd-to-6th leading orders and extend the expansion up to four higher orders, 
thus providing the correct {\it ten} leading orders of the eigenvalue for large real frequency.
We also provide the first few coefficients in the large, real-frequency expansion of the eigenfunctions, thus going, for the first time, beyond  leading
order. We compare our asymptotic expansions for both the eigenvalues and eigenfunctions with high-precision numerical calculations and find excellent agreement.
The results of this paper together with those in~\cite{Casals:2004zq} 
thus 
provide a correct, high-order asymptotic expansion of the 
 eigenvalues and eigenfunctions for large, real frequency.

The layout of the rest of the paper is as follows.
In Sec.~\ref{sec:eq} we introduce the spin-weighted spheroidal equation and its symmetries.
In Sec.~\ref{sec:asympts} we perform the large frequency asymptotic analysis of the spin-weighted spheroidal eigenfunctions and eigenvalues.
We compare our asymptotic analysis and our numerical results in Sec.~\ref{sec:num}.
In Appendix \ref{sec:coeffs} we give explicit expressions for the coefficients in the series for the  eigenfunctions and in
Appendix \ref{sec:toolkit} we describe the implementation of the asymptotics for the eigenvalue in a \textit{Mathematica} toolkit.

\section{Spin-Weighted Spheroidal Equation}\label{sec:eq}

Teukolsky~\cite{teukolsky1972rotating,Teukolsky:1973ha} managed to decouple and separate by variables the linear spin-field perturbations
of Kerr black holes.
He achieved this for the radiative components of the massless fields of spin\footnote{The symbol $s$ really corresponds to the helicity of the spin-field, although in keeping with general convention, we refer to it as the spin.} $s=0$ (scalar), $\pm 1/2$ (neutrino), $\pm 1$ (electromagnetic) and $\pm 2$ (gravitational).
The polar-angular factors of the perturbations are the so-called spin-weighted spheroidal harmonics ${}_{\indhel}S_{\indlmf}$.
These functions satisfy the following linear, second-order ordinary differential equation (ODE):
\begin{align} \label{eq:ang. teuk. eq.}
\left(
\frac{\mathrm{d}\phantom{x}}{\mathrm{d}x}\left((1-x^{2})
\frac{\mathrm{d}\phantom{x}}{\mathrm{d}x}\right)+c^{2}x^{2}-2\indhel
cx-\frac{(m+\indhel
x)^{2}}{1-x^{2}}+\E-\indhel^2 \right)
{}_{\indhel}S_{\indlmf}(x)=0,
\end{align}
where $x\equiv \cos\theta\in [-1,+1]$ is the physical region of interest and $\theta\in [0,\pi]$ is the (Boyer-Lindquist) polar angle.
Here, $c\equiv a\omega$, where $a\in\mathbb{R}$ is the angular momentum per unit mass of the black hole and $\omega\in\mathbb{C}$ is the frequency of the field mode.
The multipole number $\ell=|s|,|s|+1,|s|+2,\dots$ serves to label the eigenvalue $\E$ and $m=-\ell,-\ell+1,\ell+2,\dots,+\ell$ is the azimuthal number. 
The ODE \eqref{eq:ang. teuk. eq.} has two regular singular points at $x=\pm 1$ and an irregular singular point at 
$x=\infty$. The eigenvalue $\E$ is chosen so that the corresponding solution ${}_{\indhel}S_{\indlmf}$ is regular over $x\in [-1,+1]$.
The case $s=0$ yields the (scalar) spheroidal equation~\cite{Erdelyi:1953,Flammer,meixner2013mathieusche}, whereas the case $c=0$ yields the  spin-weighted {\it spherical} equation~\cite{goldberg1967spin} (in which case\footnote{This corrects a typographical error in Eq.~(1.3)~\cite{Casals:2004zq}.},
$\Ecz=\ell(\ell+1)$).

Other common parameterizations of the eigenvalue are
\begin{subequations}
\begin{align} \label{eq:lambda}
{}_{\indhel}\lambda_{\indlmf} & \equiv
\E-\indhel (\indhel+1)+c^{2}-2mc,
\\
\A & \equiv 
\E-\indhel (\indhel+1).		\label{eq:A_from_E}
\end{align}

\end{subequations}
The manifest symmetries of the spin-weighted spheroidal equation imply that
\begin{align}
\label{eq:sym}
{}_{-{\indhel}}S_{\ell mc}(x)=(-1)^{\ell+m}{}_{\indhel}S_{\ell mc}(-x),\qquad 
{}_{{\indhel}}S_{\ell (-m)(-c)}(x)=(-1)^{\ell+\indhel}{}_{\indhel}S_{\ell mc}(-x),\qquad
\end{align}
where the choice of signs ensures consistency with the so-called Teukolsky-Starobinsky identities~\cite{teukolsky1974perturbations,Chandrasekhar},
and
\begin{align}
\label{eq:sym eigenvalues}
{}_{-{\indhel}}E_{\ell mc}={}_{\indhel}E_{\ell mc},\qquad
{}_{{\indhel}}E_{\ell m(-c)}={}_{\indhel}E_{\ell (-m)c}.
\end{align}
While ${}_{\indhel}\lambda_{\indlmf}$ is most common in the current literature, we shall use $\A$ in the following sections to ease comparison
with BRW.

\section{Large real-frequency asymptotics}\label{sec:asympts}

In this paper we are interested in the large, real 
``frequency" (by which we really mean $c\to \pm\infty$) behaviour of the eigenvalues and eigenfunctions.  In addition, by Eqs.~\eqref{eq:sym} and \eqref{eq:sym eigenvalues} we may assume
$c$ positive and deduce the $c$ negative behaviour from changing $m$ to $-m$.  
We therefore restrict ourselves
to $c>0$ from now on. 

We here generalize to arbitrary spin Flammer's~\cite{Flammer} approach in the scalar case -- this is essentially BRW's path, although they 
obtained some incorrect results which we specify and correct below.
We start by writing
solutions of the spin-weighted spheroidal equation \eqref{eq:ang. teuk. eq.} as
\begin{align} \label{eq:asympt. S for x->+/-1}
{}_{\indhel}S^{\pm}_{\indlmf}(x)=
(1-x)^{|m+s|/2}(1+x)^{|m-s|/2}
e^{-c(1\mp x)}\g_\pm(x),
\end{align}
where $\g_\pm(x)$ are regular functions. The powers of $(1-x)$ and $(1+x)$ are dictated by the Frobenius method, so that the  solution
 is regular at both boundary points $x=+1$ and $-1$. The exponential factor
 $e^{-c(1\mp x)}$ is included for convenience when looking for an asymptotic solution ``near" $x=\pm 1$.
 
From Eq.~\eqref{eq:sym}, it follows that the  solution $g_-(x)$  is obtained from
$g_+(x)$ under the transformation $\{s\to -s, x\to -x\}$, modulo an overall sign of the solution.
Hence, from now on we focus on $g_+$.

Looking for the asymptotic solution valid near $x=+1$,  we introduce  $u\equiv 2c(1-x)$.  (For a full discussion see \cite{Casals:2004zq}, where this procedure defines an asymptotic solution\footnote{The function  $S^{\text{inn},+1}(x)$ in  \cite{Casals:2004zq} corresponds to the leading-order term in the expansion for $g_+$ which we provide in this paper.} $S^{\text{inn},+1}(x)$ for $0\leq 1-x \leq  \mathcal{O}(c^\delta)$ with $-1< \delta < 0$.)
Inserting the expression \eqref{eq:asympt. S for x->+/-1} into  the ODE \eqref{eq:ang. teuk. eq.}, we find that $\g_{+}(u)$ satisfies the following equation,
\begin{align}\label{eq:ODE g}
&u\g_{+}''+\left(|m+s|+1-u\right)\g_{+}'+\tfrac{1}{2}\left(\q-|m+s|-1-s\right)\g_{+}+\nonumber\\
&\qquad \As\ \g_{+}-
\frac{1}{4c}\Bigl(u\left(u\g_{+}''+\left(|m+s|+|m-s|+2-u\right) \g_{+}'\right)+\nonumber\\
&\qquad\qquad\tfrac{1}{2}\bigl((|m+s|+1)(|m-s|+1)-(m+1)^2+s^2  -  \bigl(|m+s|+|m-s|+2 + 2s\bigr) u \bigr)\g_{+}\Bigr)=0,
\end{align}
where primes denote derivatives with respect to $u$.
Here we have defined
\begin{align}
	\As\equiv \frac{1}{4c}\left(s(s+1)-m(m+1)+c^2-2\q c+\A\right),
\end{align}
introducing the  parameter $\q$, discussed in the Introduction, which is chosen  so that 
\begin{align}\label{eq:exp As}
	\As=o\left(1\right), \qquad \text{as\ }c\to \infty .
\end{align}

BRW left the parameter $\q$ undetermined for non-zero spin. Its value may be determined by requiring that the number of zeros of our asymptotic expansion\footnote{Specifically, this number is obtained by adding the number of zeros of the leading order expression Eq.~\eqref{eq:1F1} near $x=+1$, the number of zeros of its counterpart near $x=-1$ and the number of zeros $z_0$ in Eq.~\eqref{eq:z_0} near $x=0$.} is equal to the number of zeros of the spin-weighted spheroidal harmonics (which is given in Eq.~(4.1) in Ref.~\cite{Casals:2004zq}). Ref.~\cite{Casals:2004zq} determined this value to be\footnote{It can be checked that our expressions for $\q$
and $z_{0}$ here are equivalent to --but simpler than-- those given in Eqs.~(4.5) and (4.6) in~\cite{Casals:2004zq}.}
\begin{align}
	\q &= \ell+1-z_{0},		    \quad && \text{if} \quad \ell \geq {}_{|s|}\ell_{m},   \\
	\q &= 2\ell + 1 - {}_{|s|}\ell_{m}, \quad &&  \text{if} \quad \ell < {}_{|s|}\ell_{m},
\end{align}
where 
${}_s\ell_{m}\equiv |m+s |+s$,
and
\begin{align}\label{eq:z_0}
	z_{0}\equiv 
	\begin{cases}
		0 & \text{if} \quad \ell+m \quad \text{even},   \\
		1 & \text{if} \quad \ell+m \quad \text{odd}.
	\end{cases}
\end{align}
The value of $z_0$ indicates whether ${}_{\indhel}S_{\indlmf}(x)$ has a zero ``near" $x=0$ for large-$c$: it is $z_{0}=1$ if ${}_{\indhel}S_{\indlmf}(x)$ has a zero
``near" $x=0$ and it is $z_{0}=0$ if it does not.
Regarding the values of $\q$, we note, in particular, that: (i) $\q$ is an integer if $s$ is an integer, whereas $\q$ is a half-integer if $s$ is a half-integer;
 (ii) ${}_{\indhel}q_{\ell m}={}_{-\indhel}q_{\ell m}$.

In the limit of infinitely large $c$, only the first line in Eq.~(\ref{eq:ODE g}) survives and  the solution of the resulting ODE
which is regular at $u=0$ $(x=1)$ is
\begin{align}\label{eq:1F1}
{}_1F_{1} (-{}_sp_{\ell m+},|m+s|+1,u),
\end{align}
where we have introduced ${}_sp_{\ell m+}\equiv \frac{1}{2}\left(\q-|m+s|-s-1\right)$ and  ${}_1F_{1}$ is the regular confluent hypergeometric function~\cite{Erdelyi:1953}.
We note that ${}_sp_{\ell m+}\in\mathbb{Z}$ for $2s\in\mathbb{Z}$.

Eq.~\eqref{eq:ODE g} then suggests that we express the function $\g_{+}$ as 
\begin{align}\label{eq:g Laguerre}
\g_{+}(u)=\sum_{n=-\infty}^{\infty}\an{n}\  {}_1F_{1}(-{}_sp_{\ell m+}-n,|m+s|+1,u),
\end{align}
where without loss of generality we assume that $a_0=1$.
The series coefficients $\an{n}$ satisfy a three-term recurrence relation
\begin{align}
\label{eq:rec rln}
 &(2 n + \q  - |m - \indhel|+ \indhel+1) (2 n + \q  -  |m + \indhel|- \indhel+1) a_{n+1} + \nonumber\\
 &\qquad 2  \bigl( 8 c n   - (2 n+\q)^2  +  2\indhel^2 -(m+1)^2 - 8 c \As\bigl)a_{n}+\nonumber\\
 &\qquad\qquad (2 n + \q  + |m - \indhel|+ \indhel-1) (2 n + \q  +    |m + \indhel|- \indhel-1) a_{n-1} =0 .
\end{align}
These  recurrence relations are obtained by inserting the series representation \eqref{eq:g Laguerre} into \eqref{eq:ODE g} and using the following recurrence relations satisfied by the hypergeometric functions~\cite{Erdelyi:1953}:
\begin{subequations}
\begin{gather}\label{eq:L rec rln}
u\, {}_1F_1{}'' (\alpha,\beta,u) +(\gamma-u){}_1F_1{}' (\alpha,\beta,u)=
(\gamma-\beta){}_1F_1{}'(\alpha,\beta,u)+\alpha {}_1F_1(\alpha,\beta,u),\\
u\, {}_1F_1{}'(\alpha,\beta,u) =
\alpha\bigl({}_1F_1(\alpha+1,\beta,u){}-{}_1F_1(\alpha,\beta,u)\bigr),\\
u\,  {}_1F_1(\alpha,\beta,u) =\alpha {}_1F_1(\alpha+1,\beta,u)-(2\alpha-\beta) {}_1F_1(\alpha,\beta,u) +(\alpha-\beta) {}_1F_1(\alpha-1,\beta,u),
\end{gather}
\end{subequations}
for constant $\alpha$, $\beta$ and $\gamma$.
Because of the analyticity of the coefficients of the spin-weighted spheroidal differential equation in the parameter $c$, we can assume a series expansion in powers of $1/c$ for $\As$~(see, e.g., theorems 2.9 and 4.9 in Ch.~8 in Ref.~\cite{kato1966perturbation}):
\begin{align}\label{eq:exp As}
\As\sim \sum_{k=1}^\infty \Ask{k} c^{-k}
\qquad \text{as\ }c\to \infty .
\end{align}
 Correspondingly we now expand the coefficients for large-$c$:
\begin{align}\label{eq:exp an}
\an{n}\sim \sum_{k=|n|}^\infty \ank{n}{k} c^{-k}
\qquad \text{as\ }c\to \infty ,
\end{align}
 where the structure of this coefficient expansion follows from the dominant first term in the coefficient of $a_n$ in 
the recurrence relation.
We give  explicit expressions  for the series coefficients $\ank{n}{k}$ for $n:-3\to 3$ and $k: |n|\to 3$ in Appendix \ref{sec:eigenfunction coeffs}.

Inserting the expansions Eqs.~\eqref{eq:exp As} and \eqref{eq:exp an}   into the recurrence relation \eqref{eq:rec rln} 
and requiring  it to be satisfied order-by-order determines the expansion coefficients.
Specifically, we find $\A$ takes the form
\begin{align} \label{eq:series E for large w}
\begin{aligned}
	\A 	&=	-c^{2}+2\q c-\frac{1}{2}\left(\q^2-m^2+2\indhel+1\right) +\sum_{k=1}^7\frac{A_k}{c^k} +O\left(\frac{1}{c^{8}}\right).
\end{aligned}
\end{align}
Other common definitions of the eigenvalue are easily computed from this using Eqs.~\eqref{eq:lambda} and \eqref{eq:A_from_E}. Dropping the subscripts on $\q$ for compactness, the $A_k$'s are given by
\renewcommand{\q}{q}
\begin{align}
A_1=-\frac{1}{8}\left(\q^3-m^2\q+\q-2\indhel^2\left(\q+m\right)\right),
\end{align}
\begin{align}
A_2=
\frac{1}{64} \left(-m^4+6 m^2 \q^2+2 m^2-5 \q^4-10 \q^2-1+4 \indhel^2 \left( m^2+4 m \q+3 \q^2+1\right)\right),
\end{align}
\begin{align}
A_3&=
\frac{1}{512} \Bigl(-q \bigl(37 + 13 m^4 + 114 q^2 + 33 q^4 - 2 m^2 (25 + 23 q^2)\bigr) + 
\nonumber\\
 &\qquad 4 (13 m - m^3 + 25 q + 9 m^2 q + 33 m q^2 + 23 q^3) \indhel^2 - 
 8 (m + q) \indhel^4\Bigr) ,
\end{align}
\begin{align}
A_4&=
\frac{1}{1024} \Bigl(-14 + 2 m^6 - 239 \q^2 - 340 \q^4 - 63 \q^6 - 3 m^4 (6 + 13 \q^2) + 
 10 m^2 (3 + 23 \q^2 + 10 \q^4) +
\nonumber\\
&\qquad
 4 \bigl(-m^4 - 9 m^3 \q + 5 m^2 (2 + 3 \q^2) + m \q (93 + 73 \q^2) +    5 (3 + 23 \q^2 + 10 \q^4)\bigr) \indhel^2 
\nonumber\\
&\qquad
- 8 (2 + 3 m^2 + 9 m \q + 6 \q^2) \indhel^4\Bigr),
\end{align}
\begin{align}
A_5&=
\frac{1}{8192} \Bigl(-\q \bigl(1009 - 53 m^6 + 5221 \q^2 + 4139 \q^4 + 527 \q^6 + 
    5 m^4 (127 + 93 \q^2) -
\nonumber\\
&\qquad
m^2 (1591 + 3750 \q^2 + 939 \q^4)\bigr) + 
 2 \bigl(14 m^5 - 45 m^4 \q + 130 m^2 \q (3 + \q^2) - 
\nonumber\\
&\qquad
20 m^3 (7 + 18 \q^2) +  2 m (303 + 1820 \q^2 + 685 \q^4) + 
    \q (1591 + 3750 \q^2 + 939 \q^4)\bigr) \indhel^2 - 
\nonumber\\
&\qquad
 80 (7 m + m^3 + 11 \q + 9 m^2 \q + 18 m \q^2 + 10 \q^3) \indhel^4 + 
 16 (m + \q) \indhel^6\Bigr),
\end{align}
\begin{align}
A_6&=
\frac{1}{131072} \Bigl(-3747 - 51 m^8 - 86940 \q^2 - 205898 \q^4 - 101836 \q^6 - 9387 \q^8 + 
 12 m^6 (85 + 167 \q^2) - 
\nonumber\\
&\qquad
6 m^4 (939 + 5078 \q^2 + 1855 \q^4) + 
 12 m^2 (701 + 8657 \q^2 + 9575 \q^4 + 1547 \q^6) + 
\nonumber\\
&\qquad
 8 \bigl(19 m^6 + 270 m^5 \q - m^4 (191 + 309 \q^2) - 
    4 m^3 \q (919 + 725 \q^2) + m^2 (949 + 1482 \q^2 - 255 \q^4) +
\nonumber\\
&\qquad 
    2 m \q (8135 + 15310 \q^2 + 3363 \q^4) + 
    3 (701 + 8657 \q^2 + 9575 \q^4 + 1547 \q^6)\bigr) \indhel^2 + 
\nonumber\\
&\qquad
 16 \bigl(-467 + 17 m^4 - 236 m^3 \q - 3438 \q^2 - 1455 q^4 - 
    4 m \q (919 + 725 \q^2) 
- 2 m^2 (407 + 849 \q^2)\bigr) \indhel^4 + 
\nonumber\\
&\qquad
 128 (4 + 7 m^2 + 19 m \q + 12 \q^2) \indhel^6\Bigr).
\end{align}
\begin{align}
A_7	&=\frac{1}{2097152} \Bigl(-\q \bigl(822221 + 4093 m^8 + 5771940 \q^2 + 7568470 \q^4 + 2520820 \q^6 + 175045 \q^8 - \nonumber		\\
&\qquad 1540 m^6 (65 + 43 \q^2) +42 m^4 (16371 + 29350 \q^2 + 6375 \q^4) - \nonumber\\	
&\qquad4 m^2 (353449 + 1345421 \q^2 + 847819 \q^4 + 95167 \q^6)\bigr) - 
 \nonumber\\
&\qquad 
8 \bigl(257 m^7 - 1253 m^6 \q + 35 m^4 \q (379 + 169 \q^2) - 
    35 m^5 (181 + 381 \q^2) + 
	\nonumber\\&\qquad7 m^2 \q (-6821 + 6070 \q^2 + 3567 \q^4) + 
\nonumber\\
&\qquad 
    7 m^3 (5389 + 32190 \q^2 + 12045 \q^4) - 
    \q (353449 + 1345421 \q^2 + 847819 \q^4 + 95167 \q^6) - 
\nonumber\\
&\qquad 
    m (112285 + 1057707 \q^2 + 953715 \q^4 + 136773 \q^6)\bigr) s^2 + 
 \nonumber\\&\qquad112 \bigl(31 m^5 + 363 m^4 \q - 6 m^3 (107 + 131 \q^2) - 
\nonumber\\
&\qquad 
    10 m^2 \q (1135 + 749 \q^2) - \q (10573 + 23530 \q^2 + 5673 \q^4) - 
    m (5389 + 32190 \q^2 + 12045 \q^4)\bigr) s^4 + 
\nonumber\\
&\qquad 
 896 (73 m + 19 m^3 + 105 \q + 113 m^2 \q + 189 m \q^2 + 95 \q^3) s^6 - 
 640 (m + \q) s^8\Bigr).
\end{align}
\renewcommand{\q}{{}_{\indhel}q_{\ell m}}
While for compactness we have given just the first ten orders (to order $1/c^7$) for $\A$ in Eq.~\eqref{eq:series E for large w}, the process is easy to automate as it is for the $a_n$'s. We have implemented code into the SpinWeightedSpheroidalHarmonics package of the Black Hole Perturbation Toolkit to compute the high-frequency expansion of the eigenvalue -- see Appendix \ref{apdx:implementation_in_Toolkit}. We also provide additional code to compute the $a_{n,k}$'s and $A_k$'s to arbitrary order.

We note that BRW gave an expansion for $\A$ to the first six orders (i.e., to order $1/c^3$) but, while their first three orders were as in Eq.~\eqref{eq:series E for large w}, our values
of $A_1$, $A_2$ and $A_3$  correct the corresponding last three orders in Eq.~(4.12) in BRW\footnote{Eq.~(4.12) in BRW was merely
reproduced in~\cite{Casals:2004zq} and in~\cite{Berti:2005gp} without previously checking it, and so containing the last three erroneous terms of the original BRW version.}. We also note that for $s=0$ our results for the $A_k$'s agree with Ref.~\cite{Do-Nhat_2001}. Finally, we note that it could also be interesting to consider the limit where both $c\rightarrow\infty$ and $m\rightarrow\infty$ with a fixed $m/c$ ratio. This has been analysed in the $s=0$ case \cite{Hod:2015cqa} but has not, to the best of our knowledge, been analysed in the $s \neq 0$ case. We leave such an analysis for future work.

\section{Comparison with numerical calculation}\label{sec:num}

We validate our high-frequency asymptotic expansions by comparing them against a numerical calculation. For the numerical results we use the SpinWeightedSpheroidalHarmonics \textit{Mathematica} package which is part of the Black Hole Perturbation Toolkit \cite{BHPToolkit}. This package employs both a spectral method \cite{Hughes:1999bq} and Leaver's method \cite{Leaver:1985,Leaver:1986a}, combining them in a similar fashion to the method used by Ref.~\cite{Falloon_et_al} for the $s=0$ case, to rapidly compute high precision values for the spin-weighted spheroidal-harmonics and their eigenvalues. 

For the eigenvalue calculation we use the \texttt{SpinWeightedSpheroidalEigenvalue} to compare against Eq.~\eqref{eq:series E for large w}. Note that Eq.~\eqref{eq:series E for large w} gives the expansion for $\A$, whereas the \texttt{SpinWeightedSpheroidalEigenvalue} command returns ${}_{\indhel}\lambda_{\indlmf}$ so we use Eqs.~\eqref{eq:lambda} and \eqref{eq:A_from_E} to convert between them. The results of the comparison are shown in Fig.~\ref{fig:s=2,l=2,m=2} which shows that our high-frequency expansion agrees extremely well with the numerical results for large $c$. The comparison is further discussed in the figure's caption.
\begin{figure}[h!]
\begin{center}
   \includegraphics[width=0.485\textwidth]{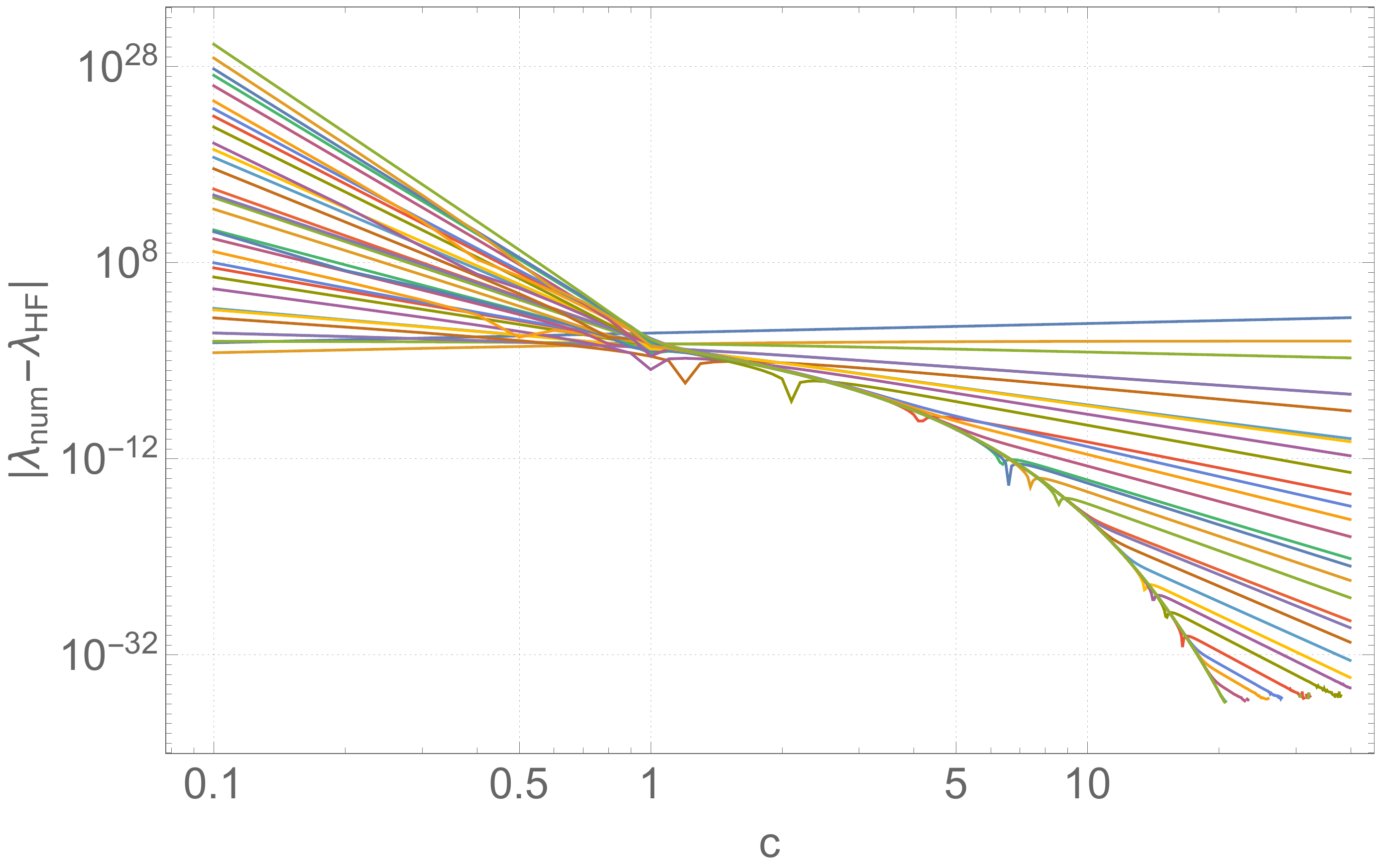}\quad
   \includegraphics[width=0.485\textwidth]{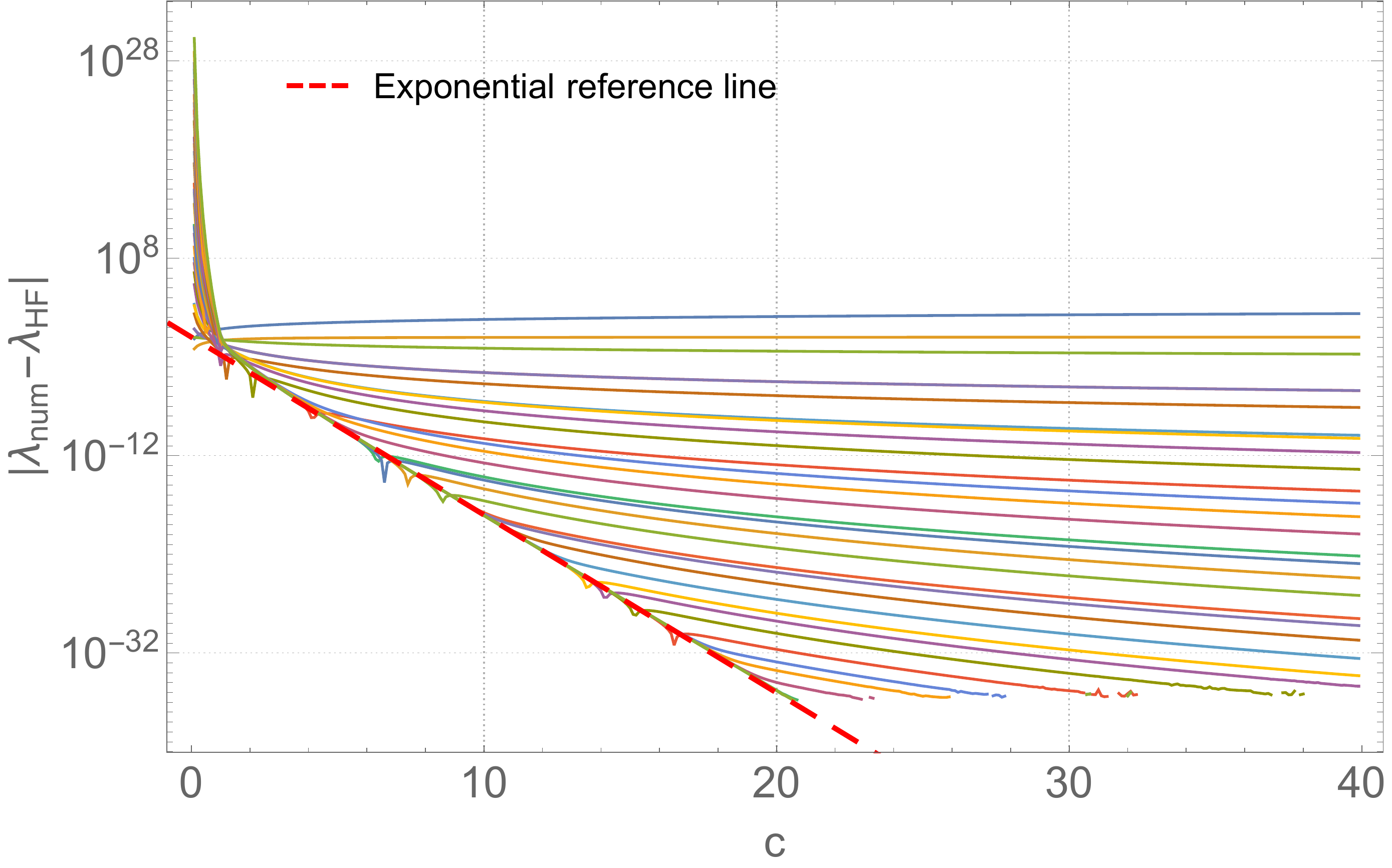}
   \end{center}
\caption{
Difference between the numerically computed eigenvalue, $\lambda_\text{num}$, (computed to 40-digits of accuracy) and its high-frequency expansion, $\lambda_\text{HF}$, for the case $\{s,\ell,m\}=\{2,2,2\}$. We present the same results on both a $\log$-$\log$ scale (left panel) and a $\log$ scale (right panel). The different curves are computed using successively higher orders in the high frequency expansion. On the right of the graph, the top curve plots the numerical value of the eigenvalue. The subsequent lower curves are computed by subtracting the high-frequency series truncated at $\mathcal{O}(c^1), \mathcal{O}(c^0), \mathcal{O}(c^{-1})\dots\mathcal{O}(c^{-27})$, respectively. For $c \gtrsim 1$ including additional terms in the high frequency series improves the comparison with the numerical results up to a point. After this, adding more terms does not improve the agreement. The shape of the curve beyond which adding terms not does improve the agreement is clearest on a $\log$ scale (right panel). This suggests that in addition to admitting a series expansion in $c^{-1}$ there is an exponential term which the power law expansion cannot capture. Finally, as one would expect of a high frequency expansion, for $c \lesssim 1$ adding terms acts to worsen the agreement with the numerical results. The eigenvalue is better approximated by small frequency expansions around $c=0$ in this region.}
 \label{fig:s=2,l=2,m=2}
\end{figure}

\begin{figure}
	\includegraphics[width=0.49\textwidth]{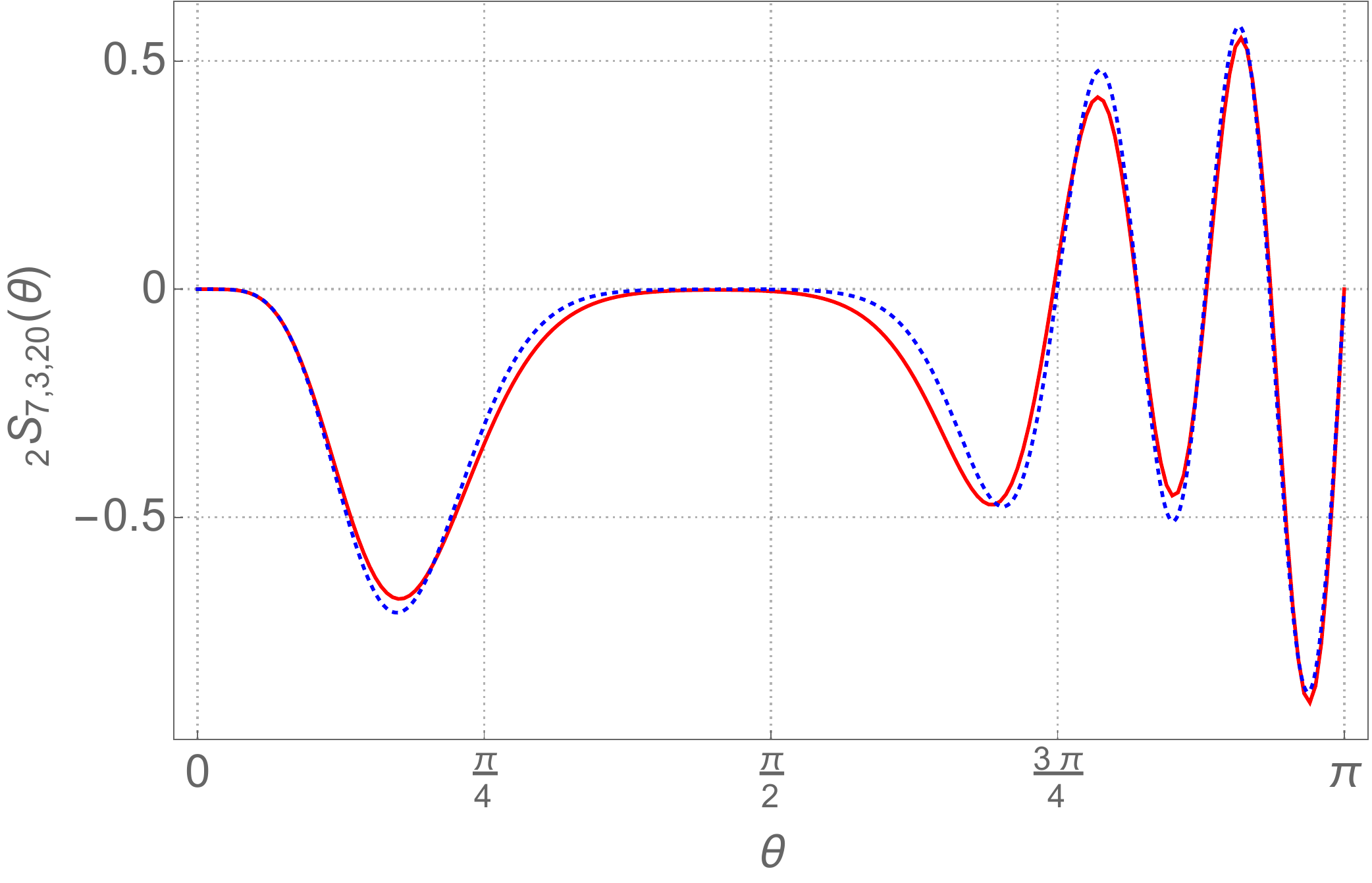}\quad
	\includegraphics[width=0.49\textwidth]{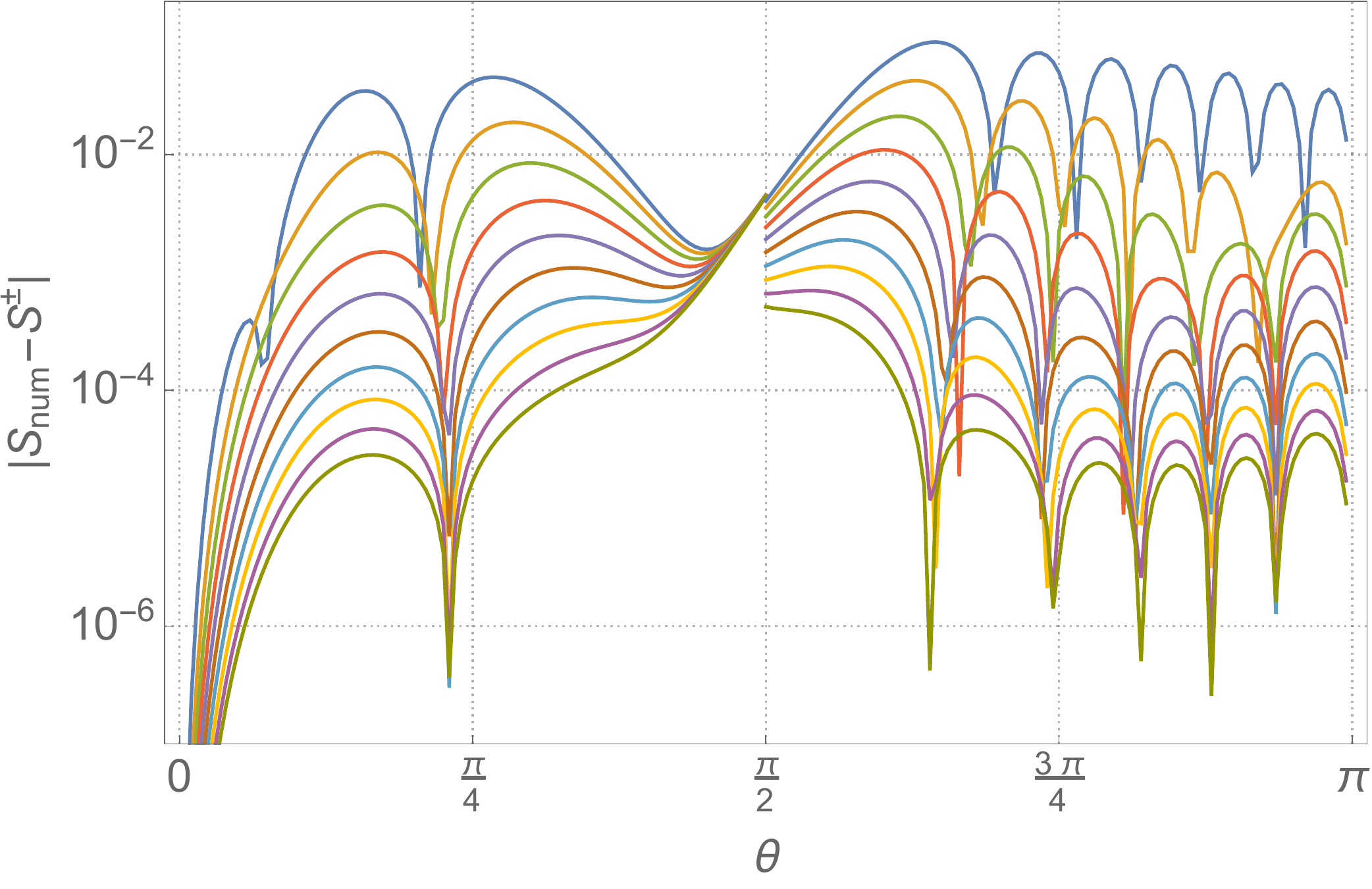}
	\caption{Example of the high-frequency approximation to the spheroidal-harmonic eigenfunction for parameters $\{s,\ell,m,c\}=\{2,7,3,20\}$. (Left panel) The (red) solid curve shows the numerically computed value of ${}_2S_{7,3,20}$. The leading-order approximation is shown with the (blue) dotted curve. (Right panel) Including higher-order terms in the expansion improves the agreement with the numerical results. In this figure the top curve is the difference between the leading-order expansion and the numerical data. Successive lower curves are the difference between the numerical expansion and successively higher-order expansions.}\label{fig:S_num_vs_approx}
\end{figure}

For the numerical calculation of the eigenfunctions we use the \texttt{SpinWeightedSpheroidal}-\texttt{HarmonicS} command which computes ${}_sS_{\ell m c}e^{im\varphi}$ in our notation. These harmonics are normalized such that
\begin{align}
	\int_0^{2\pi}\int_0^\pi {}_sS_{\ell_1 m_1 c}(\theta) e^{im_1\varphi} {}_sS_{\ell_2 m_2 c}(\theta)e^{-im_2\varphi} \sin\theta\,d\theta\,d\varphi = \delta_{\ell_1}^{\ell_2} \delta_{m_1}^{m_2},
\end{align}
where $\delta$ is the Kronecker delta function. On the other hand, we do not know the normalization of the ${}_s S_{\ell mc}^\pm$ 
in Eq.~\eqref{eq:asympt. S for x->+/-1} with the  $g_\pm$  given by \eqref{eq:g Laguerre} and its  $\{s\to -s, x\to -x\}$ counterpart.
 To make a meaningful comparison with the numerical calculation of the harmonics we numerically integrate ${}_s S_{\ell mc}^+$ over $x\in [1,0]$, and ${}_s S_{\ell mc}^-$ over $x\in [0,-1]$ to obtain their normalization. With this information we can ensure that the numerical and asymptotic approximate solutions are normalized the same. Figure \ref{fig:S_num_vs_approx} presents an example of the excellent agreement we find between the numerical calculation and the high-frequency approximation of the eigenfunctions. The convergence of the expansion in Eq.~\eqref{eq:g Laguerre} becomes slower the further the point is from $x=+1$; similarly for $g_-$ from $x=-1$. This means that the combined asymptotic expansion of ${}_sS^+_{\ell mc}$ and ${}_sS^-_{\ell mc}$ converges more slowly near x=0, as reflected in Fig.~\ref{fig:S_num_vs_approx}. The convergence near $x=0$ could be improved by incorporating the `outer' solution of Eq.~(3.26) in Ref.\cite{Casals:2004zq} in the manner done in Sec.~III.~C of that paper.

For the eigenvalues and the eigenfunctions, the excellent agreement we observe between the high-frequency asymptotics and the numerical results gives us confidence in both.

\dataccess{Code to compute the expansions in this paper to arbitrary order has been integrated into the open source Black Hole Perturbation Toolkit (\href{http://bhptoolkit.org/}{bhptoolkit.org}) -- see Appendix \ref{sec:toolkit} for more details.}
\aucontribute{MC and ACO calculated the high-order large-frequency expansions presented in this work. NW made detailed comparisons of these expansions with high-precision numerical calculations and integrated all three authors' codes into the Black Hole Perturbation Toolkit.}
\competing{We have no competing interests.}
\ethics{There are no ethical concerns regarding this work.}
\funding{MC acknowledges partial financial support by CNPq (Brazil), process number 310200/2017-2. NW gratefully acknowledges support from a Royal Society - Science Foundation Ireland University Research Fellowship. }
\ack{This work makes use of the Black Hole Perturbation Toolkit.}

\appendix

\section{Eigenfunction coefficients}\label{sec:eigenfunction coeffs}\label{sec:coeffs}

For completeness we here give the first three orders for the coefficients in the eigenfunction asymptotic expansion -- see Eqs.\eqref{eq:g Laguerre} and \eqref{eq:exp an}.
The coefficients $\ank{n}{k}$ may conveniently be expressed in terms of ${}_sp_{\ell m\pm}\equiv \frac{1}{2}\left(\q-|m\pm s|\mp s-1\right)$,
again dropping the subscripts on ${}_sp_{\ell m\pm}$ and $\q$ for compactness:
\begin{subequations}
\begin{align}
\ank{-1}{1}&=\tfrac{1}{4} p_{-}p_{+} ,\\
\ank{1}{1}&=-\tfrac{1}{4} (p_{-}  - q - s)(p_{+} -q + s ) ,
\end{align}
\end{subequations}
\begin{subequations}
\begin{align}
\ank{-2}{2}&=
\tfrac{1}{32}  p_{-} (p_{-} - 1) p_{+}(p_{+} - 1),\\
\ank{-1}{2}&=
\tfrac{1}{8}p_{-}p_{+}(q-1),\\
\ank{1}{2}&=
-\tfrac{1}{8} (p_{-}  - q - s)(p_{+} -q + s ) (q+ 1) ,\\
\ank{2}{2}&=
\tfrac{1}{32} (p_{-} -q - s )(p_{-} -q - s-1 ) (p_{+} -q + s ) (p_{+} -q + s-1 ),
\end{align}
\end{subequations}
\begin{subequations}
\begin{align}
\ank{-3}{3} =& \tfrac{1}{384} p_{-} (p_{-} - 1) (p_{-} - 2) p_{+}(p_{+} - 1)(p_{+} - 2),\\
\ank{-2}{3} =& \tfrac{1}{64}p_{-} (p_{-} - 1) p_{+}(p_{+} - 1)(2q-3) ,\\
\ank{-1}{3} =&\tfrac{1}{128}p_{-}p_{+} \bigl((p_{-}   (p_{-}  - q - s+1) - q-s -2) \times \nonumber \\
	 		 & (p_{+} (p_{+} - q + s+1) - q+s-2) + 2 (q-2) (5q-2) - 2 s^2\bigr) ,\\
\ank{1}{3}  =& \tfrac{1}{128} (p_{-} -q-s)(p_{+} -q +s)\times \nonumber \\
   			 & \bigl((p_{-}  (p_{-} - q - s+1)-2)(p_{+}  (p_{+} - q + s+1) -2)  + 2 (q+2 ) (5 q+2) - 2s^2\bigr),\\
\ank{2}{3}  =& \tfrac{1}{64}(p_{-} -q-s) (p_{-} - q-s - 1) (p_{+} -q +s)(p_{+} -q+s- 1)(2q+3),\\
\ank{3}{3}  =& -\tfrac{1}{384} (p_{-} -q - s )(p_{-} -q - s-1 )(p_{-} -q - s-2 ) (p_{+} -q + s ) 	\times   \nonumber \\
			 & (p_{+} -q + s-1 )(p_{+} -q + s-2 ) .
\end{align}
\end{subequations}

These series coefficients were not given  in BRW or, to the best of our knowledge, anywhere else in the literature.

As noted in the body of the paper, it is ${}_sp_{\ell m+}\in\mathbb{Z}$ for $2s\in\mathbb{Z}$, and it
is straightforward to show that ${}_sp_{\ell m-}\geq0$ for $s\geq0$ and ${}_sp_{\ell m+}\geq0$ for $s\leq0$.
The structure of the expanded recursion relations then shows that the functional expansion 
Eq.~(\ref{eq:g Laguerre}) terminates with finite lower limit ``$-{}_sp_{\ell m-}$" for $s\geq0$, reflected in the vanishing of the coefficients $\ank{n}{k}$ for $n<-{}_sp_{\ell m-}$. Corresponding comments hold for $s\leq0$ with ${}_sp_{\ell m-}$ replaced by ${}_sp_{\ell m+}$. For $s=0$, ${}_0p_{\ell m-}={}_0p_{\ell m+}$  and our observation agrees with Eq.~(8.2.9) of Flammer~\cite{Flammer}.

\section{Implementation in the Black Hole Perturbation Toolkit}\label{apdx:implementation_in_Toolkit}\label{sec:toolkit}

We have implemented the calculation of the high frequency expansion of the spin-weighted spheroidal eigenvalue and eigenfunction into the \textit{Mathematica} SpinWeightedSpheroidalHarmonics package which is part of the open-source Black Hole Perturbation Toolkit. This package allows for the numerical and (where possible) analytic calculation of the eigenvalue and eigenfunction of the spin-weighted spheroidal equation. It also allows the user to compute small frequency expansions of these functions using the standard Mathematica \texttt{Series[..]} function. Following this work, we have implemented the high (real) frequency expansion of the eigenfunction as well.

As an example, the high-frequency expansion of the eigenvalue, ${}_{\indhel}\lambda_{\indlmf}$, for $\{s,\ell,m\}=\{2,7,3\}$ about $c=\infty$ can be computed via
\begin{align}
	&\texttt{Series[SpinWeightedSpheroidalEigenvalue[2,7,3,c], \{c,}\infty\texttt{,5\}]} =	\nonumber \\ 
	&\qquad10 c-30-\frac{45}{c}-\frac{405}{2 c^2}-\frac{9855}{8 c^3}-\frac{17685}{2 c^4}-\frac{2261115}{32c^5}+\mathcal{O}[c^{-6}]
\end{align}
The expansion can also be computed around $c=-\infty$, for example:
\begin{align}
	&\texttt{Series[SpinWeightedSpheroidalEigenvalue[2,7,3,-c], \{c,}\infty\texttt{,5\}]} =	\nonumber \\ 
	&\qquad22 c-30-\frac{51}{c}-\frac{501}{2 c^2}-\frac{13017}{8 c^3}-\frac{24603}{2 c^4}-\frac{3283149}{32 c^5}+\mathcal{O}[c^{-6}]
\end{align}

We have also included an example notebook in the Toolkit which demonstrates the use of this function and provides code to calculate the $a_{n,k}$ and $A_k$ coefficients that appear in Eq.~\eqref{eq:exp an} and Eq.~\eqref{eq:series E for large w}, respectively.

\bibliographystyle{RS}

\end{document}